\documentclass[12pt]{article}
\usepackage{epsfig}
\def\be{\begin{eqnarray}}\def\ba{\begin{eqnarray}}
\def\ee{\end{eqnarray}}\def\ea{\end{eqnarray}}
\def\ben{\begin{enumerate}}\def\bitem{\begin{itemize}}
\def\een{\end{enumerate}}\def\eitem{\end{itemize}}

\def\roughly#1{\mathrel{\raise.3ex\hbox{$#1$\kern-.75em%
\lower1ex\hbox{$\sim$}}}}

\def\A0{A_0}
\def\bq{\begin{equation}}
\def\eq{\end{equation}}
\def\la{\langle}\def\ra{\rangle}
\def\K0{K^0}

\begin{document}

\begin{titlepage}
\begin{flushright}
KIAS-P07035
\end{flushright}

\begin{center}

 \vskip 1.5cm

{\Large \bf  Holographic Nuclear Matter in AdS/QCD} \vskip 1. cm
  {\large Youngman Kim$^{(a)}$\footnote{e-mail:~chunboo81@kias.re.kr.},
Chang-Hwan Lee$^{(b)}$\footnote{e-mail:~clee@pusan.ac.kr.}, Ho-Ung Yee
    $^{(a)}$\footnote{e-mail:~ho-ung.yee@kias.re.kr.} }

\vskip 0.5cm
(a)~{\it School of Physics, Korea Institute for Advanced Study , Seoul
  130-722, Korea}

(b)~{\it Department of Physics, Pusan National University, Busan 609-735, Korea}

\end{center}

\centerline{(\today) }
\vskip 1cm
\vspace{1.0cm plus 0.5cm minus 0.5cm}

\begin{abstract}

We study the physics with finite nuclear density in the framework
of AdS/QCD with holographic baryon field included. Based on a mean
field type approach, we introduce the nucleon density as a
bi-fermion condensate of the lowest mode of the baryon field and
calculate the density dependence of the chiral condensate and the
nucleon mass. We observe that the
chiral condensate as well as the mass of nucleon decrease with
increasing nuclear density. We also consider the mass splitting of
charged vector mesons in iso-spin asymmetric nuclear matter.

\end{abstract}

\end{titlepage}

\newpage

\section{Introduction}

Inspired by AdS/CFT correspondence\cite{Maldacena:1997re}, several attempts to construct
a holographic model of QCD appeared recently, in both bottom-up\cite{EKSS,PR,Evans:2005ip}
and top-down\cite{sakai-sugimoto} approaches. The morale is to introduce a dynamically
generated additional space, which roughly corresponds to the
energy scale of the field theory, and try to construct a
holographic dual model that captures important aspects of the
original 4D field theory, such as large N expansion, confinement,
and chiral symmetry breaking et cetera. The amazing observation is that these non-perturbative aspects
of strongly coupled field theory, which are highly non-trivial to analyze,
can be described in simple terms in the 5D holographic dual model,
sometimes called
AdS/QCD model\footnote{We will later refer AdS/QCD model for phenomenological bottom-up constructions only, but
for now we mean both top-down and bottom-up.}.
In a sense, the holographic model may be viewed as a relevant effective theory
for strongly coupled, non-perturbative field theory.
The counter-intuitive aspect at first sight is that this effective theory
necessitates a new space dimension generated dynamically. Note that this
does not necessarily mean that the additional dimension is physical; in view of the 4D field theory,
it is simply a derived notion that facilitates a magically simple description of
important aspects of strongly coupled field theory.

The origin of the simplicity in the 5D dual model is traced back to
the large N limit as was the case in the original AdS/CFT correspondence in string theory.
We should thus keep in mind that a usual simple, tree-level calculation in the holographic dual model
is capturing the leading N contributions, and we are bound to suffer from sub-leading
corrections.\footnote{There is no framework in phenomenological AdS/QCD models yet to do a systematic $1\over N$ expansion.}
Within this limitation, the AdS/QCD model may be taken as a new alternative
effective theory of low energy QCD.

The simplest bottom-up construction for AdS/QCD\cite{EKSS,PR},
sometimes called the hard-wall model, takes a slice of 5D AdS
spacetime\footnote{Our metric convention is $\eta_{MN}={\rm
diag}(+----)$, and we denote 5D coordinates by capital letters
while 4D coordinates are written with Greek letters. We will
 follow mostly the convention in Ref.\cite{EKSS}.}
\be ds^2 = {1\over z^2}\left(-dz^2 + dx^\mu dx^\nu
\eta_{\mu\nu}\right)\quad, \ee with $0 \le z \le z_m$, where $z_m$
is the IR cut-off corresponding to the confinement scale, which
must be fitted to physical observables later. In this paper, we
will concern only with $N_F=2$, (u,d) sector of QCD, whose
extension to $N_F=3$, (u,d,s) sector will be left open to a future
work. Corresponding to the global chiral symmetry $SU(2)_L \times SU(2)_R$ of
QCD, we introduce local gauge fields $A_L$ and $A_R$ in the 5D AdS
slice, whose values at $z=0$ act as external sources for $SU(2)_L$
and $SU(2)_R$ currents respectively. More precisely,
$(A_L)_\mu^{ij}|_{z=0}$ couples to $(j^\mu_L)^{ij}=\bar q_L^i
\gamma^\mu q_L^j$, and similarly for $SU(2)_R$ sector. The chiral
symmetry breaking, explicitly by the small current quark mass
$(M_q)_{ij}\bar q_L^i q_R^j+{\rm h.c.}$, as well as spontaneously
by the chiral condensate $\Sigma^{ij}=\left<\bar q_R^j
q_L^i\right>\ne 0$ is holographically realized by a VEV of a
scalar field $X$ which is bi-fundamental with respect to the gauge
group $SU(2)_L \times SU(2)_R$, \be \left<X(z)\right>= {1\over
2}M_q z +{1\over 2}\Sigma z^3=\left({1\over 2} m_q z+ {1\over
2}\sigma z^3\right) {\bf 1}\quad, \ee where in the last equality,
we assume iso-spin symmetry for simplicity\footnote{Iso-spin
violating effects from $m_u\ne m_d$ are neglected in this work.}.
From the standard 5D action for the field $X$, \be S_X=\int dz
\int dx^4 \sqrt{G_5}\,{\rm Tr}\left( |DX|^2 +3|X|^2\right)\quad,
\ee with $DX=\partial X-i A_L X + i X A_R$, the above VEV breaks
$SU(2)_L\times SU(2)_R$ to the diagonal iso-spin $SU(2)_I$ as
expected. The tower of (pseudo)vector and scalar mesons, including
pions, arise from the normalizable KK modes of the above 5D
fields. In addition to the parameters appearing in the above, the
5D $SU(2)_L\times SU(2)_R$ gauge coupling constant in \be S_g=\int
dz \int dx^4 \sqrt{G_5}\,{\rm Tr}\left( -{1\over 4
g_5^2}F_{MN}F^{MN}\right)\quad, \ee specifies the model completely
for the two-flavor meson sector. After fitting these parameters,
the physical observables are predicted, which compare well with
experiments\cite{EKSS,PR}.

We point out that this phenomenological model in fact encodes essential ingredients
of the top-down model of Sakai-Sugimoto\cite{sakai-sugimoto}, in that the relevant 5D part of the 10D spacetime
is approximated by an AdS slice and the world volume gauge fields on two branches of D8 branes
are identified with $A_L$ and $A_R$ respectively, whose breaking to the diagonal $SU(N_F)_I$
is achieved by tachyon condensation $\left<X\right>\ne 0$ that joins the two D8 branes.

Since the nucleons, i.e. the protons and neutrons,
 are integral objects in QCD, it is pertinent to include them in the
AdS/QCD model. An early attempt to see the high spin versus mass Regge trajectory was pioneered
by Brodsky-Teramond\cite{deTeramond:2005su}, but recently Ref.\cite{Ho}
focused on the lowest spin $1\over 2$ nucleons
with chiral symmetry breaking effects carefully included, and provided us with a concrete
 AdS/QCD model with holographic nucleons.
This will be our framework for studying the physics of finite nuclear density.

We comment that the model in Ref.\cite{Ho} again mirrors the salient aspects of the top-down description of spin
$1\over 2$ baryons in the Sakai-Sugimoto model \cite{HRYY}. Spin $1\over 2$ baryons from quantized small
instanton-solitons on the D8 brane \cite{Son:2003et,Nawa:2006gv,Hata:2007mb}
 are effectively described
by a 5D Dirac spinor field of $SU(2)_F$ fundamental. As we identify two branches of D8 brane as representing
separate $SU(2)_L$ and $SU(2)_R$ sectors in the phenomenological AdS/QCD model (which are combined to $SU(2)_I$
at the IR), the 5D Dirac spinor on the two branches of D8 brane will corresponds to two separate 5D Dirac
spinors in the AdS/QCD model, which are fundamental under $SU(2)_L$ and $SU(2)_R$ respectively. Indeed, this is
the field content of the model in Ref.\cite{Ho}.

Chiral symmetry is believed to be restored at high density  and/or temperature, such as in the core of neutron
stars or in the early stage of  fire balls in the relativistic heavy ion collisions.
Since the discovery of the pulsar \cite{Hulse75}, neutron stars have provided extreme environments to test the
validity of general relativity, especially the gravitational wave radiation \cite{Burgay03} and the emission of
short hard gamma ray bursts \cite{Nakar07} from neutron star mergers. Many observatories to detect the
gravitational waves (such as LIGO, GEO, and Virgo) and gamma-ray bursts (such as Swift and HETE II) from neutron
star mergers are in operation or under development. Neutron stars can also set-up nuclear matter of extreme
density in which chiral symmetry is restored. Various phase transitions, such as kaon condensation \cite{CHLee},
are suggested in the dense core of neutron stars \cite{Brown06,Latt07}. Even though the core of neutron star
cannot be observed directly, the maximum mass of neutron star which can be observed strongly depends on the
equation of state of neutron star \cite{Latt07}. The change in the maximum mass of neutron stars results in the
change in the ratio of neutron star-black hole binaries to double neutron star binaries, which can be detected
by the gravitational wave observation \cite{Bethe07,Kalogera07}. In addition, the inner structure of neutron
star will be revealed by a detailed pattern of the detected gravitational wave radiation from the gravitational
wave observatories. Hence  theoretical investigation on the physics of extreme dense matter is very important in
order to understand the physics of the neutron stars.

At high temperature, in addition to the chiral phase transition, QCD deconfinement phase transition from
hadronic matter to the perturbative (weakly interacting due to asymptotic freedom) quark gluon plasma (QGP) is
expected. Such QGP states might have been formed at the early stages of the evolution of our Universe after Big
Bang. However, experimental results of the recent relativistic heavy collision (RHIC) indicate that the matter
after deconfinement (and chiral) phase transition  is nothing like a weakly interacting perturvative QGP
\cite{Gyu05,Shu05}. Instead, the matter formed after phase transition is a strongly interacting quark gluon
liquid. These results initiated very active research on the physics of hot and dense matter after chiral and
deconfinement phase transition \cite{SZ04,Asa03,Pet04,PLB05}.

Since the properties of nuclear matter cannot be calculated from the fundamental QCD Lagrangian due to strong
coupling, various effective theories, such as chiral perturbation theory, have been introduced to treat nuclear
matter phenomenologically. Most of the parameters in these approaches were fixed by the experimental values at
low densities (from vacuum to normal nuclear matter density), at which chiral symmetry is spontaneously broken.
Hence, the validity of the extrapolation of the effective theories, which were built in the symmetry broken
phase, to high densities where chiral symmetry is restored, may be questioned \cite{Brown06}. In this respect,
recent holographic dual AdS/QCD approach which is based on the chiral symmetry at its construction provides a
new tool to investigate the physics in the symmetry restored phase.

In this work, we introduce a new way of describing nuclear matter in the framework of
AdS/QCD model with holographic baryon fields.
 We turn on the nuclear density in mean field approach by a non-zero 5D
 bi-nucleon condensate, and this subsequently affects various other holographic fields,
 such as $X$ and $U(1)_B$ gauge field $V$, to result in several physical consequences that we
 study in this paper. 
 We point out the relation between our approach and the recent approaches
by baryon chemical potential\cite{Kim:2006gp,Nakamura:2006xk,Kobayashi:2006sb,Domokos:2007kt}.
In the spirit of AdS/CFT correspondence,
the quark chemical potential in 4D QCD is encoded in the boundary value of the
5D $U(1)_B$ bulk gauge field, ~$\mu_q\psi^\dagger\psi\leftrightarrow V_0(x,z)=\mu_q+\cdots$~.
This is quite similar to a way of introducing various chemical potentials in chiral perturbation theory,
where the global chiral symmetry is promoted to a local gauge one, and a chemical potential is
introduced as the time component of the gauge potential~\cite{KST, SSCh}.
%
The physics behind this approach can be thought of as follows.
When we introduce finite density of baryons as instanton-solitons, they are energetically
pulled down towards IR boundary and deposit there. From the Chern-Simons coupling,
these IR localized instanton-charge density will source the time-like component of the 5D $U(1)_B$ gauge field $V_0$,
and the vector potential profile of non-zero baryon chemical potential results.
A hidden assumption in this picture is that the IR-deposited baryons have a $\delta$-function localized
wavefunction at the IR boundary, and we solve free EOM for $V_0$ in the bulk
without describing IR-localized baryons explicitly. Although this is correct as a first approximation,
our analysis based on the spread nucleon profile along z-direction can be viewed as
a more precise analysis of the situation.
We also comment that the interaction terms of $V_0$ and meson
fields, say $ {\cal H}$, feature the commutator structure, ${\cal L}_{\rm int}\sim [V_0, {\cal H}]$, and
therefore the chemical potential $\mu_q$ will  be inoperative to neutral mesons,
with an exception of Chern-Simons couplings studied in
~\cite{Domokos:2007kt}.

After briefly reviewing some details about our AdS/QCD model with
holographic baryon fields in the next section,
we evaluate the density dependence of the chiral condensate and the nucleon mass.
We also calculate the mass splitting of charged vector mesons in the iso-spin asymmetric nuclear matter.
Finally,
we  re-analyze the in-medium nucleon mass in a more phenomenological approach, where  we use the density dependence of
the chiral condensate
determined
through Hellmann-Feynman theorem and the Gell-Mann-Oakes-Renner relation\cite{qC}.
 In this case, we find
the nucleon mass drops drastically with increasing density up to the normal nuclear matter
number density $\rho_0 (\sim 0.16 ~fm^{-3})$.\\

\section{The model with holographic baryons}

In this section, we briefly summarize the model in Ref.\cite{Ho}
to set-up the stage.
For simplicity, for $N_F=2$ meson sector we take the simplest hard-wall AdS/QCD model
that was explained in the Introduction.
To describe spin $1\over 2$, iso-spin $1\over 2$ baryons
in this set-up, with the nucleons (proton and neutron) as the lowest
iso-spin doublet, we introduce 5D (Dirac) spinor fields $N$ corresponding to
the 4D baryon operators $\cal O$ with the same spin and charges in the original QCD.
Note that once we introduce a field in the 5D dual model, we will actually get
a tower of 4D states as we can equivalently describe the system by Kaluza-Klein reduction along
our compact extra dimension. The idea behind this is that
these spin $1\over 2$, iso-spin $1\over 2$ 4D operators $\cal O$ in QCD can
create/annihilate the whole spectrum of 4D baryons $B_n$ with the same quantum numbers,
\be
\left<B_n|{\cal O}|0\right> \ne 0 \quad, \,\,n=0,1,2,\cdots\quad,
\ee
and in the holographic dual model, these $B_n$'s are packed into a single 5D spinor field $N$ (of the same quantum numbers)
as the normalizable Kaluza-Klein modes along the extra dimension.
This is why holography maps a 5D field $N$ to a 4D operator $\cal O$,
not to a specific 4D state such as nucleons. Therefore,
any holographic model is bound to include the excited spectrum as well, in addition to the lowest
state proton and neutron we are interested in.

Looking at the problem more carefully, one may be puzzled by a few things.
In the AdS/QCD model side, we have two gauge fields $A_L$ and $A_R$ for $SU(2)_L\times SU(2)_R$
chiral symmetry which are broken to the iso-spin $SU(2)_I$ by a Higgs mechanism with $X$.
Any field must have a specific representation under $SU(2)_L\times SU(2)_R$.
In QCD side, nucleons (and their excitations) form a doublet under $SU(2)_I$
and since they are states in the broken chiral symmetry phase, it is not at all obvious
to even talk about their $SU(2)_L\times SU(2)_R$ quantum numbers.

A guide to way-out may be seen in thinking about
how the chiral symmetry breaking is realized
in the AdS/QCD side.
The symmetry $SU(2)_L\times SU(2)_R$ exists as a definition of the theory, and its
breaking is achieved by putting a VEV of $X$; this allows a theoretical or fictitious limit of
turning off the VEV of X restoring the chiral symmetry.
In QCD side, this would again be an imaginary limit of chiral symmetry restoration,
and every operator will have a definite charge under $SU(2)_L\times SU(2)_R$.
This allows us to
match the quantum numbers of 5D fields to the quantum numbers of 4D operators.
We then analyze how chiral symmetry breaking affects these 4D operators, and
in the 5D AdS/QCD side these must correspond to turning on the VEV of the scalar $X$.
Therefore, the effects from chiral symmetry breaking
in the AdS/QCD model should be realized by possible gauge invariant couplings to the field $X$.

Although
whether or not the above presumptuous
chiral symmetry restoration can actually be realized is unclear,
it may be taken as simply an intermediate tool to identify 5D field contents and their couplings to
the Higgs field $X$.
This chiral symmetry restoration limit was in fact considered by 't Hooft
in his arguing that the massless chiral nucleon doublet should exist in this limit to match
UV anomalies of $SU(2)_L\times SU(2)_R$ with those in IR.
More precisely, a left-handed chiral doublet ${\cal O}_L=(p_L,n_L)^T$ of fundamental representation under $SU(2)_L$,
and similarly a right-handed chiral doublet ${\cal O}_R=(p_R,n_R)^T$ of $SU(2)_R$
exist in this limit, and the chiral symmetry breaking to the diagonal $SU(2)_I$
induces a mass coupling $\bar{\cal O}_L {\cal O}_R +{\rm h.c.}$ that results in the observed
nucleon mass $m_N\sim 0.94$ GeV.
Hence, we are led to introduce two 5D spinor fields $N_1$ and $N_2$
whose charges under $SU(2)_L\times SU(2)_R$
are $({1\over 2},0)$ and  $(0,{1\over 2})$, corresponding to the 4D operator ${\cal O}_L$ and ${\cal O}_R$
respectively.\footnote{By ${\cal O}_{L,R}$, we mean the QCD operators $(q_Lq_Lq_L)$ and $(q_R q_R q_R)$
that create massless chiral nucleon states,
with a slight abuse of notation compared with the above which means the massless states themselves.}
Moreover, we have to ensure that in the chiral symmetry restoration limit $\left<X\right>=0$,
we must have a massless left-handed zero mode from $N_1$ and a massless right-handed zero mode
from $N_2$ in their Kaluza-Klein reduction. They are the massless chiral nucleon doublets to match
chiral anomalies in the restoration limit. This requirement in fact unambiguously fixes the IR boundary
conditions for $N_1$ and $N_2$.

The chiral symmetry breaking by $\left<X\right>\ne 0$ must then introduce a mass coupling between
$N_1$ and $N_2$. The relevant gauge invariant coupling of the lowest dimension is
\be
{\cal L}_m = -g \bar N_1 X N_2 +{\rm h.c.}\quad,
\ee
with a strength $g$ that must be fitted to reproduce the nucleon mass $m_N=0.94$ GeV as the lowest
mass eigenvalue of the KK spectrum.
Note that the would-be massless chiral zero modes from $N_1$ and $N_2$
are lifted to the single massive nucleon state with $m_N=0.94$ GeV, while the mixing
between already massive excitations from
separate $N_1$ and $N_2$ will split them into a
parity doubling pattern of massive excitations.
Here, the definition of 4D parity involves the exchange of $N_1$ and $N_2$, so that
the two nearly degenerate states split by mixing $N_1$ and $N_2$ have opposite 4D parity.
The parity doubling for excited baryons is thus a prediction of the model.

One last thing to mention is the chirality of the 4D QCD operators ${\cal O}_L$ and ${\cal O}_R$
and its relation to the 5D Dirac spinors $N_1$ and $N_2$.
One may be puzzled about this since 5D spinor doesn't have a chirality.
The resolution lies in the fact that in 5D the signature of the
Dirac mass term flips its sign under parity.
The magnitude of the 5D Dirac mass $m_5$ for $N_1$ and $N_2$ is fixed by an AdS/CFT relation\cite{Henningson:1998cd,Contino:2004vy}
\be
(m_5)^2=(\Delta -2)^2\quad,
\ee
with $\Delta$ being the scaling dimension of ${\cal O}_{L,R}$, which we take
the free theory value $\Delta={9\over 2}$ for simplicity.\footnote{We believe
considering anomalous dimension to $\Delta$ would be an important refinement from
the analysis in Ref.\cite{Ho}.}
With a positive $m_5$ for a 5D spinor $N$,
it can be seen that only the right-handed component $N_R$ of $N$
survives kinematically near the AdS boundary $z=0$,
while the left-handed component decays fast enough to become a normalizable mode.\footnote{Our convention
is $i\gamma^5 \psi_L=+\psi_L$ and the mass term is ${\cal L}_m=-m_5 \bar N N$.}
The non-normalizable $N_R$ near the boundary then couples to a left-handed chiral operator ${\cal O}_L$
in QCD side by $\bar N_R {\cal O}_L +{\rm h.c.}$. Note that this is consistent with the fact that
$N$ and ${\cal O}_L$ have the same quantum number, and the possible normalizable chiral zero mode
from $N$ is left-handed matching to the chirality of ${\cal O}_L$.
With a negative $m_5$, the story is simply reversed.\footnote{See also Ref.\cite{Contino:2004vy} for
the case when both chiral components are normalizable with a specific range of $m_5$.}
Therefore, the right choice for our model is $m_5={5\over 2}$ for $N_1$ and $m_5=-{5\over 2}$ for $N_2$ respectively.

In summary, the AdS/QCD model of spin ${1\over2}$, iso-spin ${1\over 2}$ baryons is
given by the action
\be
S_{\rm kin}& =& \int dz \int dx^4 \sqrt{G_5}\,\left[i \bar N_1 \Gamma^M D_M N_1+i \bar N_2 \Gamma^M D_M N_2
-{5\over 2}\bar N_1 N_1
+{5\over 2}\bar N_2 N_2\right]\quad,\nonumber\\
S_m &=& \int dz \int dx^4 \sqrt{G_5}\,\left[ -g \bar N_1 X N_2 -g \bar N_2 X^\dagger N_1\right]\quad,\label{action}
\ee
where the covariant derivatives for $N_1$ and $N_2$ include the gauge group $SU(2)_L \times SU(2)_R$
as well as the metric connection, and a single parameter $g$ should be fixed to reproduce the lowest mass
nucleon mass $m_N=0.94$ GeV. By expanding $N_1$ and $N_2$ in terms of KK eigenmodes, it is easy to find the
eigenmode equations that must be solved to find the mass spectrum of 4D spin $1\over 2$ baryons.
Writing $N_1(x,z)=f_{1L}(z)B_L(x)+f_{1R}(z)B_R(x)$ and similarly
for $N_2(x,z)=f_{2L}(z)B_L(x)+f_{2R}(z)B_R(x)$, where $B_{L,R}$ are
the components of the 4D eigenmode spinor $B=(B_L, B_R)^T$ with mass $m_N$ to be determined, we have
\begin{eqnarray}
&&\left(
\begin{array}{cc}
\partial_z -{\Delta \over z} & -{g \left<X\right>\over z} \\
-{g \left<X^\dagger\right>\over z} & \partial_z -{4-\Delta \over z}
\end{array}
\right)
\left(
\begin{array}{c} f_{1L} \\ f_{2L}
\end{array}
\right)=-m_N
\left(
\begin{array}{c} f_{1R} \\ f_{2R}
\end{array}
\right)\,, \nonumber\\
&&\left(
\begin{array}{cc}
\partial_z -{4-\Delta \over z} &{g \left<X\right>\over z}\\
{g \left<X^\dagger\right>\over z} & \partial_z -{\Delta \over z}
\end{array}
\right)
\left(
\begin{array}{c} f_{1R} \\ f_{2R}
\end{array}
\right)=m_N
\left(
\begin{array}{c} f_{1L} \\ f_{2L}
\end{array}
\right)\,,
\label{kk_mode}
\end{eqnarray}
with $\Delta$ ($={9\over 2}$ in our case) in general.
As mentioned before, the existence of the chiral zero modes when $\left<X\right>=0$ requires us
the IR boundary condition  $f_{1R}(z_m)=f_{2L}(z_m)=0$.
From the meson sector, Ref.\cite{EKSS,PR} found the best fit for $\left<X\right>={1\over 2}(m_q z+\sigma z^3)$
with $m_q=2.34$ MeV and $\sigma=(311 \,{\rm MeV})^3$, as well as the IR cut-off $z_m=(330\, {\rm MeV})^{-1}$.
Then, the only remaining parameter of the theory is the dimensionless coupling $g$,
which was found to be $g=9.3$ in Ref.\cite{Ho} to reproduce $m_N=0.94$ GeV as a lowest mass eigenvalue.

\section{Finite nuclear density in AdS/QCD}
\label{sec2}
The baryon chemical potential, introduced as a background
for the time component of the bulk $U(1)_B$ gauge
field~\cite{Kim:2006gp,Nakamura:2006xk,Kobayashi:2006sb,Domokos:2007kt},
is of limited use for some observables
due to a specific commutator structure of the interaction terms.
An example would be its effect to the chiral condensate which is encoded in
the VEV of the scalar $X$ because $DX=\partial X-i[V,X]$.
In the present work, we
introduce the nuclear density through the mean field of the 4D nucleon-bilinear, {\it i.e.},
$\rho_s=\la\bar\psi(x)\psi (x)\ra$, where $\psi(x)$ is the 4D nucleon field, and
$\rho_s$ is the iso-scalar baryon number density.~\footnote{Note that
the iso-scalar density $\rho_s$ is roughly equivalent to the baryon number density $\rho_B=\la \psi^\dagger (x) \psi(x)\ra$
at low density~\cite{CHLee}: $\rho_s\approx \rho_B -\la \frac{p^2}{m_N^2}\psi^\dagger\psi\ra$, where $p$ is the
baryon momentum. } From the term ${\cal L}_m=-g \bar N_1 X N_2 +{\rm h.c.}$, we
can discuss its effects on the chiral condensate, et cetera.

\subsection{In-medium chiral condensate}

Without nuclear density, the Yukawa coupling ${\cal L}_m$ in Eq.~(\ref{action}) is
a purely cubic interaction term, and
it does not enter the equation of motion for the bulk scalar field $X$ or its VEV,  $\la X\ra\equiv X_0$.
 As is well-known, $X_0$ has the following profile
\ba
X_0(z)=(\frac{1}{2} m_q z+\frac{1}{2}\sigma_0 z^3){\bf 1}\, ,
\ea
where $m_q$ is the current quark mass and $\sigma_0\sim (0.330~{\rm GeV})^3$ is the chiral condensate in the vacuum.
For simplicity, we take the chiral limit, $m_q=0$, as our result is expected not to change much from
a small quark mass.
As we turn on a finite nuclear matter and introduce the non-zero mean field
$\left<\bar N_1 N_2+\bar N_2 N_1\right>$  of the baryon bi-linear,
\ba
\left<\bar N_2 N_1 +\bar N_1N_2\right>=(f_{2R}^2-f_{1R}^2)\left<\bar\psi (x) \psi(x)\right>
\Longrightarrow (f_{2R}^2-f_{1R}^2)\rho_s\, ,
\ea
where  $N_1=f_{1L}(z)\psi_L(x)+f_{1R}(z)\psi_R(x)$, $N_2=f_{2L}(z)\psi_L(x)+f_{2R}(z)\psi_R(x)$,
and we used the important parity relation
\be
\left(\begin{array}{c} f_{1L} \\ f_{1R}\end{array}\right)=\left(\begin{array}{cc}
0 & 1\\ -1 & 0\end{array}\right)\left(\begin{array}{c} f_{2L}\\f_{2R}\end{array}\right)\quad,
\label{wave}
\ee
which is a consequence of even parity of the nucleons, the interaction (\ref{action})
modifies the equation of motion for $X$ and hence the chiral condensate.
We assume that only the lowest lying baryons, proton and neutron,
have non-zero mean field value of their bi-linears, since the higher baryon resonances cost too much energy
to form a Fermi sea. At very high density, of course, we will have to consider dense matter of higher resonances, too.
The equation of motion for $X$ is now
\ba
\left(\partial_z^2 -\frac{3}{z}\partial_z +\frac{3}{z^2}
\right) X=\frac{1}{4}\frac{g}{z^2}(f_{2R}^2-f_{1R}^2) \rho_s\, ,\label{sEoM}
\ea
and we solve it numerically with $\rho_s\sim \rho_B$ as a varying parameter.

\begin{figure}
\centerline{\epsfig{file=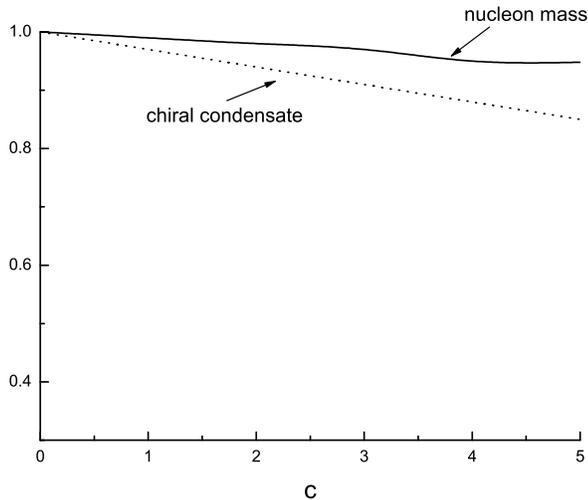,width=8.8cm}} \caption{\small
Density dependence of the nucleon mass and the chiral condensate
normalized with respect to those in the vacuum. Here $c=\rho_B/\rho_0$.}
\label{ChMn}
\end{figure}

It is an important question to specify a physically sensible IR boundary condition
for $X$ at $z=z_m$, while our chiral condensate $\sigma$ sits in the UV asymptotic $X(z)\sim {1\over 2}\sigma z^3$
as $z\to 0$ in the massless limit.\footnote{This means that
the UV boundary condition is ${X(z)\over z}\to 0$ as $z\to 0$.}
Physically, what drives the chiral condensate in the vacuum or the VEV of $X$
would be a strong QCD IR dynamics, which may be attributed to some
unspecified dynamics on the physical IR brane at $z=z_m$ in the present AdS/QCD model.
We naturally expect that this IR-localized dynamics
is not very much affected by the presence of our bulk bi-fermion condensate,
as we imposed the IR boundary condition on our baryons such that their bi-linears vanish at $z=z_m$.
Therefore, we are led to identify $X(z)$ at $z=z_m$ with its value in the vacuum,
$X(z_m)={1\over 2}\sigma_0 z_m^3$, as our IR boundary condition.
We show our numerical result for the ratio of the density dependent
chiral condensate to that  in the vacuum, $\sigma\over \sigma_0$,
  in fig.~\ref{ChMn}, where $c=\rho_B/\rho_0$
 with $\rho_0$ being the normal nuclear matter density.
We see that the chiral condensate is a decreasing function of the nuclear density.

\subsection{In-medium masses of nucleons and charged vector mesons}

In this subsection, we calculate the density dependence of the nucleon mass as well as the mass
splitting of the charged vector mesons.
We first consider the nucleon mass.
As seen in the previous subsection, the profile of
$X$ depends on the iso-scalar density, and plugging it back to our 5D  mass coupling (\ref{action}),
the nucleon wave function equation (\ref{kk_mode}) will also get a repercussion resulting
a shift in its eigenvalue nucleon mass $m_N$, not to mention its wave functions.
In fact, for a fully consistent analysis, we should solve the equation (\ref{sEoM}) for  $X$  and the equation
(\ref{kk_mode})
for the nucleon wave functions $f_{(1,2)(L,R)}$
simultaneously, as we have a cubic interaction (\ref{action}) between them.
For this purpose, we performed an iterative analysis until we get a stabilized solution. Therefore our result
is numerically reliable, and
the result is shown also in Fig.~\ref{ChMn}. We observe that the nucleon mass decreases
in the finite nuclear density.

In light of the results from other conventional
QCD effective theories, for example Ref.~\cite{BR91, Weise, KKL, qC},
both the in-medium nucleon mass and the chiral condensate have been argued to
drop by  $\sim 20\%$ at $\rho_B=\rho_0$ ($c=1$).
Our results for the density dependence from Fig.~\ref{ChMn} seem a little smaller than these previous estimates.
This might be attributed to the
fact that our analysis is based on various crude assumptions in the AdS/QCD model,
but it is also worth exploring possible improvements of the model.
One thing is that we have ignored a back-reaction on the metric due to the finite nuclear density, for example, see~\cite{GMTY, Sin07}.
Suppose we incorporated the back-reaction due to the nucleon density, then a modified metric would depend
on the nuclear density, and would introduce additional density dependence into the game other than Fig.~\ref{ChMn}.
Changing the metric would correspond to setting-up a new effective theory around the Fermi surface studied in
\cite{BR91, HKR, HY}, similar to the spirit of quasi-particles.
Instead of delving into this endeavor,
one possible phenomenological way of incorporating  this new  density dependence
is to simply accept the density dependent
chiral condensate determined from other
low energy effective theories, say in Ref.\cite{qC},
and to focus on solving the nucleon mass (\ref{kk_mode}) only. This will be discussed in the next section.

We now move on to the mass of  vector mesons at finite density.
The equation of motion for the time component of the bulk vector field turns out to be
\ba
\left ( \partial_z^2-\frac{1}{z}\partial_z   \right ) V_0^3 (z) =-g_5^2
\frac{1}{z^3} (f_{1L}^2+f_{1R}^2) (\rho_p-\rho_n)\, ,\label{EoMV0}
\ea
where $\rho_p (\rho_n)$ is  the proton (neutron) number density, and the index $3$ represents
 the iso-spin quantum number. Here we assume an iso-spin asymmetric
 environment $\rho_p\neq\rho_n$, which  in turn
induces an iso-spin chemical potential. This means that once the iso-spin density is turned on, $V_0^3$
will develop the following classical profile,  $V_0^3=c_1 +c_2z^2$ as $z\rightarrow 0$, as
a consequence of solving (\ref{EoMV0}),
where $c_1$ is the iso-spin chemical potential $\mu_I$ ($=\mu_p-\mu_n$) and $c_2$
corresponds to the iso-spin density $\rho_I$($= \rho_p-\rho_n$). Let us denote $\rho_p=x\rho_B$
and  $\rho_n=(1-x)\rho_B$.
In solving the above equation (\ref{EoMV0}), one may take two different approaches.
One would simply take the density as an input in the UV boundary condition, and
impose the Neumann boundary condition $\partial_z V_0^3=0$ at IR $z=z_m$ without
breaking the symmetry by boundary condition. This is in accord with our motivation
that a nuclear density is introduced as a bulk effect, without resorting to an unspecified
IR-localized baryon dynamics to generate $V_0$ profiles. Another approach would be
to treat both $\mu_I$ and $\rho_I$ as inputs, as we normally know the one when we are given the other.
For example, $\mu_{p,n}=\sqrt{m_N^2+(k_F^{p,n})^2}$ and $\rho_{p,n}=(k_F^{p,n})^2/(3\pi^2)$ relate
the two by $k_F^{p,n}$. In this case, we  start from the
given UV $z\to 0$ asymptotic by $\mu_I$ and $\rho_I$, and solve (\ref{EoMV0}) into the IR region.
In the present work, we adopt the latter scheme.

\begin{figure}
\centerline{\epsfig{file=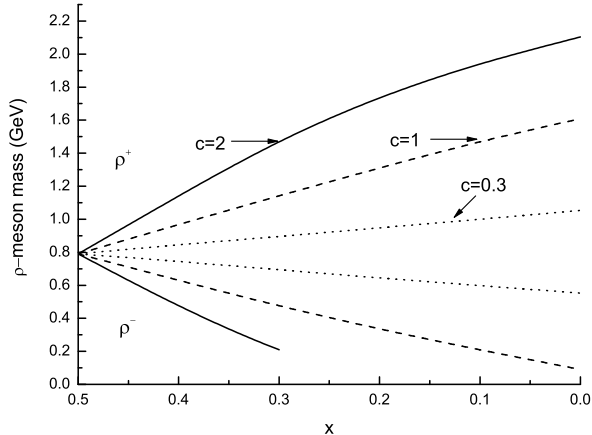,width=8.8cm}} \caption{\small
Mass splitting of charged $\rho$-mesons in iso-spin asymmetric matter ($x\neq 0.5$).
Here $x$ is the proton number density fraction, $\rho_p=x\rho_B$. Note that small $x$
corresponds to large $|\mu_I|$.}
\label{cV}
\end{figure}

 To obtain the in-medium mass of the charged vector mesons,
$V_\mu^{\pm}$, which are linear combinations of $V_\mu^1$ and $V_\mu^2$, we
consider their equation of motion in the presence of finite nuclear density.
 With a Kaluza-Klein mode expansion, $V_\mu^{\pm}(x,z)= g_5\sum_n f_n^{\pm}(z) V_\mu^{(n)} (x)$,
 it is not difficult to arrive at
 \ba
 \left ( \partial_z^2-\frac{1}{z}\partial_z +{m_n^\pm}^2\pm 2F(z) m_n^\pm + F(z)^2  \right ) f_n^\pm(z)=0\, ,
\label{rho}
 \ea
 where $F(z)$ is the solution of the previous equation (\ref{EoMV0}) for $V_0^3$.
 Here we impose the following boundary conditions: $f_n^\pm(0)=0,~\partial_z f_n^\pm(z_m)=0$
 for a normalizable eigen-function.
 We focus on the $n=1$ mode, which are the charged $\rho$-mesons, and the results are shown in Fig.~\ref{cV}.
 Note that we are considering negative iso-spin chemical potential, which is relevant for the neutron star.
 As  $|\mu_I |$ increases, the energy cost of putting additional
  particle with a negative iso-spin quantum number into the system
 will decrease, and  the mass of the particle will be reduced by $|\mu_I|$~\cite{SSCh}.
 Our results in  Fig.~\ref{cV} confirm this physics.
We remark, however, that as $x\rightarrow 0$, {\it i.e.}, as $|\mu_I |$ increases,
it is known that the pion condensation comes into the story~\cite{SSCh}, and the ground-state of the system needs to be redefined,
 and therefore our results in Fig.~\ref{cV} are no longer valid at small $x$.

\section{More phenomenological approach}
\label{sec3}

As mentioned in the previous section, density dependence of chiral condensate in our work is rather weak,
and this leads to the corresponding weak dependence of
the nucleon mass on the nuclear density.
If one is simply interested in the nucleon mass in the finite nuclear density,
it is logically meaningful to simply take the chiral condensate from
other effective theories that one might trust more, and to just focus on its effects
to the nucleon mass through the 5D mass coupling (\ref{action}).
More explicitly, we take
$X=c_2 z^3$ with $c_2=\sigma(\rho_B)/2$, where
$\sigma(\rho_B)$ is the density dependent chiral condensate
from the results of previous estimates, and we solve (\ref{kk_mode}) to re-calculate the in-medium nucleon mass.
 Another possibility may be to use the in-medium
$\rho$-meson mass, instead of the vacuum $\rho$-meson mass $\sim 770$ MeV, to fix the
density dependent IR cutoff $z_m$,
but we will not pursue this in the present work.

We adopt the model-independent
$\sigma(\rho_B)$ in Ref.\cite{qC}, which is valid at low density,\footnote{Higher order corrections to
the model independent in-medium chiral condensate are studied in Ref.\cite{LK}, and it is shown that
there are at most $15~\%$ corrections up to two times the normal nuclear matter density.  }
\ba
\sigma(\rho_B)\approx \sigma (\rho_B=0) \biggl(1-0.37\frac{\rho_B}{\rho_0} \biggr)\, .\label{qCph}
\ea
\begin{figure}
\centerline{\epsfig{file=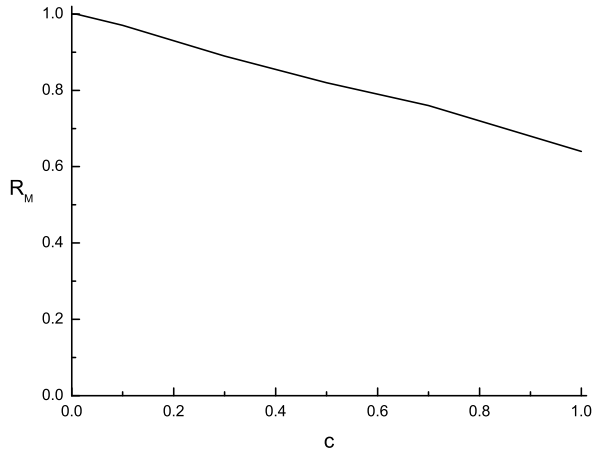,width=8.8cm}} \caption{\small In-medium nucleon mass
in the more phenomenological approach. Here $R_M\equiv M_N(\rho_B) / M_N(\rho_B=0)$.}
\label{ChMn3}
\end{figure}
and plugging (\ref{qCph})
into the KK mode equation (\ref{kk_mode}) for the nucleons, we obtain the in-medium
nucleon mass as shown in Fig.~\ref{ChMn3}. At normal nuclear matter density, it drops
about $30~\%$, which seems to agree with the results from other low energy
effective theories of QCD~\cite{BR91, Weise, KKL, qC}.
This indicates that the nucleon mass is
rather closely related to the chiral condensate.

\section{Summary}
\label{sec4}

In this paper, we study the physics of
finite density of nuclear matter in the framework of AdS/QCD with
holographic baryon fields.
Based on a mean field type approach, we turn on the nucleon density by
bi-nucleon condensates, and  calculate the density
dependence of the chiral condensate and the nucleon mass.
Our result shows that that the chiral condensate as well as the nucleon mass drop with
an increasing nucleon density.
 We also investigate a mass splitting of the charged $\rho$-mesons  in an iso-spin
 asymmetric nuclear matter, and the result is compatible with our expectation
 from a simple physical reasoning.
We finally study the in-medium nucleon mass in a more phenomenological approach and find that the
decrease of the nucleon mass in the nuclear matter is strongly
correlated to the density dependence of the chiral condensate.

It would be quite interesting to extend our work to the three flavor case including strange quarks.
Especially, the
scaling of the kaon effective mass at high density is very important for the neutron star equation of state,
because kaon is the least massive boson with a strange quantum number.
This is left to a future work.

\vskip 1cm
\section*{ Acknowledgments}

We thank Mannque Rho for useful comments, and Shin Nakamura, Piljin Yi for helpful
discussions.
C.H.L. is supported by Grant No. R01-2005-000-10334-0 (2005)
from the Basic Research Program of the Korea Science \& Engineering Foundation.
H.U.Y.
is partially supported by the Korea Research Foundation Grant (KRF-2005-070-c00030),
and thanks Kimyeong Lee for a financial support from his fund.


\end{document}